
\input phyzzx

\Pubnum={\vbox{ \hbox{CERN-TH-6464/92}\hbox{FTUAM-9209}}}
\pubnum={CERN-TH.6464/92}
\date={April, 1992}
\pubtype={}
\titlepage

\title{A PROPOSAL FOR $D>1$ STRINGS?}\foot{Talk
presented at the
Trieste Spring Workshop.  April 8-11, 1992.}
\vskip 1.0cm
\author{L. Alvarez-Gaum\'e \break J.L.F. Barb\'on
\foot{ Permanent address: Departamento
de F\'{\i}sica Te\'orica,
Universidad Aut\'onoma de Madrid,
Canto Blanco, Madrid. Spain.}
\break}

\address{Theory Division, CERN\break
 CH-1211 Geneva 23, Switzerland}

\abstract{ Using the reduced formulation
of large-$N$ Quantum Field Theories we study strings in
space-time dimensions higher than one.  Some preliminary results
concerning the possible string susceptibilities and general
properties of the model are presented.}

\endpage

\pagenumber=1

\chapter{INTRODUCTION}
In this talk I would like to report on some preliminary work
done in collaboration with J.L.F. Barb\'on on the application
of large-$N$ reduced models \REF\ek{T. Eguchi and H. Kawai,
Phys. Rev. Lett.{\bf 48} (1982) 1063.}
\REF\tek{A. Gonzalez-Arroyo and M. Okawa, Phys. Rev. {\bf D 27}
(1983) 2397, Phys. Lett. {\bf B 133}(1983) 415, Nucl. Phys.
{\bf B 247} (1984) 104.}
\REF\das{ For a review with references to the literature see
S.R. Das, Rev. Mod. Phys. {\bf 59} (1987) 235.}
[\ek,\tek,\das] to the study of non-critical
strings in dimensions higher than one.  This is a
rather challenging
problem, and at the moment we can only
make few general remarks
on a proposal to construct these  theories.  The first
lattice formulations of string theories
in arbitrary dimensions appeared
in the mid-eigthies
\def\npb{Nucl. Phys.\ }
\def\prd{Phys. Rev.\ }
\def\prl{Phys. Rev. Lett.\ }
\def\plb{Phys. Lett.\ }
\REF\adf{J. Ambj\"orn, B. Durhuus and J. Fr\"ohlich,
\npb {\bf B 259} (1985) 433.}
\REF\adfj{J. Ambj\"orn, B. Durhuus and J. Fr\"ohlich
and P. Orland, \npb {\bf B 270} (1986) 457.}
\REF\david{F. David, \npb {\bf B 257} (1985) 53,
\npb {\bf B 257} (1985) 543.}
\REF\kazakov{V.A. Kazakov, \plb {\bf 150 B} (1985) 282.}
\REF\kpz{A.M. Polyakov, Mod. Phys. Lett. {\bf A2} (1987) 893;
V. G. Knizhnik, A.M. Polyakov and A.B. Zamolodchikov,
Mod. Phys. Lett. {\bf A 3} (1988) 819.}
\REF\ddk{F. David, Mod. Phys. Lett. {\bf A 3} (1988) 1651;
J. Distler and H. Kawai, \npb {\bf B 321} (1988) 509.}
[\adf,\adfj,\david,\kazakov], and there was a good deal of
activity in the solution of some two-dimensional models
in random triangulated surfaces
\REF\models{V.A. Kazakov, \plb
{\bf 119 B} (1986) 140, D. Boulatov
and V. A. Kazakov, \plb{\bf 186 B} (1987) 379;
I.K. Kostov, and M.L. Mehta, \plb {\bf 189 B} (1987) 118;
B. Duplantier and I. Kostov, \prl {\bf 61} (1988) 1433; etc.}
[\models].  A breakthrough took place with the work in
ref. [\kpz] which provided a description in the continuum
of the coupling of minimal conformal models to two-dimensional
gravity in the light-cone gauge.  The subsequent formulation
of these models in the conformal gauge and the generalization
of the results to surfaces of arbitrary topology appeared
in [\ddk]. With the discovery of the double scaling limit
\REF\doubles{M.R. Douglas, S. Shenker,
\npb {\bf B 335} (1990) 635;
E. Br\'ezin and V.A. Kazakov, \plb {\bf 236 B} (1990) 144;
D. Gross and A.A. Migdal, \prl {\bf 64} (1990) 127,
\npb {\bf B 340} (1990) 333.}
[\doubles] we begun to understand some of the perturbative
and non-pertubative properties of non-critical strings
in dimensions below or equal to one.  Little progress has been
made however in going beyond $d=1$ apart from some evidence
that it is possible to construct Liouville Quantum Field
Theory in some especial dimensions $d=7,13,19$
\REF\gervais{See J.-L. Gervais, Comm. Math. Phys. {\bf 138}
(1991) 301; Phys. Lett. {\bf 243 B} (1990) 85,
and references therein.}.

Some interesting problems involving the presence of external
have also been studied
\REF\grossbre{E. Brezin and D. Gross, \plb {\bf  97 B}
(1980) 120.}
\REF\grossnew{D. Gross and Newman,
\plb {\bf 266 B} (1991) 291;
PUPT-1282, Dec. 91.}
\REF\wittenkon{E. Witten, Surveys in
Diff. Geom. {\bf 1} (1991)
 243, {\it On the Kontsevich Model and Other Models of
Two Dimensional Gravity}.  IAS preprint HEP-91/24.  M.
Kontsevich,
Funct. anal. i ego pril. {\bf 25} (1991) 50;
{\it Intersection
Theory on the Moduli Space of Curves and the Matrix
Airy Function}.  M. Plank Institute
for Mathematics Preprint,	October 91.}
\REF\tata{S.R. Das, A. Dhar, A.M. Sengupta and S.R. Wadia,
Mod. Phys. Lett. {\bf A5} (1990) 1041.}
[\grossbre,\grossnew,\wittenkon,\tata].  In particular
the work in [\tata] attempted to generalize the standard
one-matrix model to include the presence of local curvatures
in the triangulated surfaces described by the
 large-$N$ limit.  We will
make contact with this work later in this lecture.

The outline of this talk is as follows: In section one
 we present the problem of strings in arbitrary
dimensions in general. In section two
we briefly review the definition
of the reduced models
in the large-N description of matrix
field theories, and the general
features of the effective one-matrix model formulation of
strings propagating in $D$-dimensional flat space-time.
In section three we present the general features of the model
and study the simplest possible approximation leading to a
polymer phase. In section four we study more precise
approximations as well as the critical properties of some
effective actions suggested from our model.  We show that
for these effective actions we can obtain positive
string susceptibilities
$\gamma_{\rm st}=n/(n+m+1)$ with $n,m$
arbitrary positive integers.  We compute correlators of
macroscopic loop operators on the surface and show that only
for a subclass of critical points ($n=1$)
is it possible to define macroscopic loops in the continuum
limit.  Furthermore the continuum limit of these theories
contains an extra state with respect to the standard pure
gravity case, representing the breaking of the surface
in two pieces.
This  extra state is probably the shadow of the
tachyon in the approximations to our
model that we study.  We also explore some of
the scaling operators appearing in truncations of the
reduced model.  This section contains our main results.
The outlook and possible
future avenues to be investigated,
together with the conclusions
appear in section five.

\chapter{LARGE-$N$ REDUCED THEORIES AND MATRIX MODELS}

In the study of strings propagating in flat $D$-dimensional
space-time, we represent the sum over two-dimensional metrics
in Polyakov's approach to string theory
\REF\polyakov{A.M.Polyakov,\plb {\bf 103 B} (1981) 207, 211. }
[\polyakov] by a sum over
tringulated surfaces.   The quantity we would like to evaluate
[\adf,\david,\kazakov] is
$$
Z=\sum_g \kappa^{2(g-1)} \sum_T {e^{-\mu |T|}\over |n(T)|}
\int \prod_{i\in T_0}\sigma_i^{\alpha} d^DX_i
\prod_{\langle ij\rangle\in T_1} G(X_i-X_j)
\eqn\stringsum
$$
Where $g$ is the genus of the triangulation.  For a given
triangulation $T$, $T_0,T_1,T_2$ are respectively
the sets of vertices, edges and faces of $T$, $n(T)$ is the
order of the symmetry group of $T$, $|T|$ is the
total area of $T$ counting that every triangle has unit
area.  $\mu$ is the bare
two-dimensional cosmological constant,
$X_i$ describes the embedding of the triangulation into
$D$-dimensional flat space-time and  $G(X-Y)$ is the
propagator factor for each link.  We have included also the
local volume factor $\sigma_i ^{\alpha}$ at site $i$ to
represent the effect of local curvature. Usually we take
$\alpha=D/2$, but it is more convenient
to leave this exponent arbitrary
to include the effect of local curvature terms on the
world-sheet. For a triangulation, if $q_i$ is the local
coordination number at site $i$ (the number of triangles
sharing this vertex),
$$
\sigma_i={q_i\over 3},
$$
and the volume factor in the measure can be shown to be related
in the continuum to an expansion of the form:
$$
\sum_i \ln \sigma_i =c_0 +c_1\int \sqrt{g} R + c_2\int\sqrt{g}
R^2\ldots ,
$$
thus representing some of the effects of world-sheet curvature.  The
general features of the phase diagram in the $(D,\alpha)$ plane
were studied for example in
\REF\bkkm{D. Boulatov, V.A. Kazakov, I. Kostov, A.A. Migdal,
\npb {\bf B 275} (1986) 641.}
[\bkkm].  In the standard Polyakov formulation, the propagator
is Gaussian,
$$
G(X)=e^{-X^2/2}
$$
By standard large-N analysis the sum (2.1) can be transformed
[\david,\kazakov] into the large-$N$ expansion of a matrix
field  theory in $D$-dimensions
$$
Z=\lim_{N\rightarrow \infty} \log\int D\phi
\exp \left(-N\int d^D X Tr({1\over 2}\phi G^{-1}\phi
+{1\over 3} g\phi^3)\right)
\eqn\mastersum
$$
It is difficult to proceed very far with Gaussian propagators.  If
we work in dimensions  $D<6$ it should not matter whether we
replace the Gaussian propagator by the Feynman propagator.
There is numerical evidence that this change does not affect
the critical properties of the theory below six dimensions
\REF\boulkaz{D. Boulatov, V.A. Kazakov,
\plb {\bf 214 B} (1988) 581;
J. Ambj\"orn, Acta Phys. Pol. {\bf B21} (1990) 101,
and references therein.}
[\boulkaz].  Notice that in this construction the exponent
$\alpha$ in (2.1) is set to zero.  If we want to include the
effect of local world-sheet curvature, we can follow [\tata]
and change the kinetic term in \mastersum.  The two-dimensional
cosmological constant is here represented by $g$, $g=e^{-\mu}$.  To
summarize, we want to study the critical properties of
the action:
$$
Z=\int D\phi(X) \exp \left(-N\int d^D X Tr({1\over
2}\partial_{\mu}\phi \partial^{\mu}\phi+{g\over 3}\phi^3)\right)\ \ ,
\eqn\masterone
$$
or
$$
Z=\int D\phi(X) \exp \left(-N\int d^D X Tr({1\over 2}
A \partial_{\mu}\phi
 A\partial^{\mu}\phi+{g\over 3}\phi^3)\right)\ \ ,
\eqn\mastertwo
$$
where $A$ is a constant $N\times N$ matrix and $\phi$ is an
$N\times N$ matrix field.  The reason why the $A$-matrix in the
kinetic term simulates the effect of local curvature can be
seen by writing the propagator in \mastertwo\ explicitly
$$
\langle \phi_{ij}(X) \phi_{kl}(Y)\rangle = A^{-1}_{jk}A^{-1}_{li}
G(X-Y).
$$
If we ignore for the time being the propagator factor $G$, for
every closed index loop in a generic $\phi^3$ graph made of
$q$ propagators, we obtain a contribution of $trA^{-q}$.
Since the $\phi^3$ graphs are dual to triangulation, this
means that we are associating a curvature factor of
$trA^{-q}$ to the vertex dual to the face considered.  In this
way we can simulate the presence of the $\sigma^{\alpha}$ term
in the measure in \mastersum. We are interested in particular
in the computation of the string susceptibility exponent
$$
\chi \sim {\partial^2F\over\partial g^2}
\sim (g-g_c)^{-\gamma_{\rm st}} ,
\eqn\stringsus
$$
where $F$ is the free energy of the system, and $g_c$ is the
critical value of $g$ indicating the location of the critical
point.  Apart from the string susceptibility we would also like
to compute the spectrum of scaling operators and the properties
of the quantum geometry implied by \masterone \mastertwo.

To simplify the arguments we concentrate on the study of
planar configurations (spherical topologies).  From work dating
back to the late seventies
\REF\witten{E. Witten, {\it Recent Developments
in Gauge Theories},
Proceedings of the 1979 Carg\`ese
Summer Institute.  G. 'tHooft et al. eds.
Plenum Publ. New York.}[\witten], and due
to the factorization properties of the leading large-$N$
approximation, the planar limit is dominated by a single constant
configuration; Witten's master field (a master orbit in the case
of gauge theories).  This idea is explicitly realized in
$0$-dimensional matrix models
\REF\brezin{E. Br\'ezin C. Itzykson, G. Parisi
and J.B. Zuber, Comm. Math. Phys. {\bf 59} (1979) 35.
} [\brezin] and in
lattice gauge  theories one can think of the reduced Eguchi-Kawai
(EK)[\ek] or Twisted Eguchi-Kawai (TEK)[\tek] models as explicit
descriptions of the master field.  We now briefly describe
the main ideas behind the reduced models.  Further details
can be found in the previous
references and in the review article
[\das].  These models
were originally formulated to describe the large-$N$
properties of lattice
gauge theories.

Due to the factorization property in the
large-$N$ limit, the Schwinger-Dyson
equations for Wilson-loops form a closed system of equations
in the planar limit.  For gauge theories we get an infinite set
of polynomial equations for the Wilson-loops
\REF\loopeq{Yu. M.Makeenko, and A.A. Migdal,
\plb {\bf 88 B} (1979) 135; S. Wadia,\prd {\bf D 24} 1981) 970.  For
more details and references see
A.A. Migdal, Phys. Rep. {\bf 102} (1983) 200.}
[\loopeq].  In terms of the standard link variables
$U_{\mu}(x)$, the EK proposal consists of making them
independent of $x$, $U_{\mu}(x)\mapsto U_{\mu}$, and the
action of the reduced model becomes
$$
S_{EK}={1\over {\rm Vol.}}S(U_{\mu}(x)\mapsto U_{\mu}).
\eqn\sek
$$
Some of the properties of this action are:

1).  We obtain a theory on a single hypercube with periodic boundary
conditions.  This is an important simplification of the problem.

2).  In the case of gauge theories, the gauge symmetry becomes a global
symmetry, $U_{\mu}\mapsto \Omega U_{\mu} \Omega^{-1}$.

3).  There is an extra $U(1)^D$ symmetry in the case
of $U(N)$ lattice gauge theory $U_{\mu}\mapsto e^{i\theta_{\mu}}
U_{\mu}$.  In the case of $SU(N)$ the symmetry becomes
$Z_N^D$.

4).  It is possible to show that the planar
Schwinger-Dyson equations
following from the reduced action coincide with those obtained
from the original Wilson theory as long as open loop expectation
values vanish.  This can be shown to be true at strong coupling,
but it does not hold at weak coupling
\REF\heller{ G. Bhanot, U. Heller, H.Neuberger, \plb {\bf 113 B}
(1982) 47.} [\heller].  Without this
problem we would have a beautiful
implementation of the Master Field idea, because the
loop equations for the 1-site EK model are identical with
the standard ones.  The problem is connected with the fact
that at weak coupling the extra $U(1)^D$ symmetry is broken.
Open loops have non-trivial charge with respect to this
symmetry.  Only if the symmetry
remains unbroken are we
guaranteed to maintain the loop eqauations without extraneous
terms.

The resolution of this problem motivated the
formulation of the TEK model [\tek].  It is inspired on
'tHooft's use of twisted boundary conditions in gauge
theories
\REF\thooft{G. 'tHooft, \npb {\bf B 153} (1979) 141, Acta Phys.
Austriaca, Suppl. {\bf 22} (1980) 531.}
[\thooft].  For any matrix theory the TEK prescription
is exceedingly simple.  We reduce according to
$$
\phi(x)\mapsto D(x) \phi D(x)^{-1},
\eqn\tekred
$$
where $D(x)$ is a projective representation of the
$D$-dimensional lattice translation group.  We have
to chose a set of $D$ $N$-dimensional matrices
$\Gamma_{\mu}$ such that:
$$
D(x)=\prod_{\mu}\Gamma_{\mu}^{x_{\mu}}\qquad x=(x_1,x_2,\ldots,x_D)
\qquad x_i\in Z,
\eqn\dtrans
$$
since $D(x)$ has an adjoint action on the fields $\phi$,
the $\Gamma_{\mu}$'s are required to commute
only up to an element of the center of $SU(N)$,
$$
\Gamma_{\mu}\Gamma_{\nu}=Z_{\nu\mu}\Gamma_{\nu}\Gamma_{\mu}
$$
$$
Z_{\mu\nu}=e^{2\pi i n_{\mu\nu}/N},
\eqn\tmatrices
$$
and the integers $n_{\mu\nu}$ are defined mod $N$.  The reduced
action prescription is now
$$
S_{TEK}={1\over {\rm Vol.}}S(\phi(x)\mapsto D(x)\phi
D(x)^{-1}),
\eqn\tekaction
$$
and similarly for expectation values.

To avoid open loop expectation
values and a mismatch between the original and reduced
Schwinger-Dyson equations, the matrices $\Gamma_{\mu}$
must verify some conditions.
In particular they should
generate an irreducible representation
of the group of lattice translations.
For a lattice with $L^D$ sites this
requires $N=L^{D/2}$.  Hence
$N=L$ for $D=2$ and $N=L^2$ for $D=4$.  Hence, if we want to
simulate a two-dimensional lattice with $64\times 64$ sites
it suffices to consider the group $SU(64)$.  The choice
of twist matrix $n_{\mu\nu}$ depends on the dimensionality.
In two-dimensions, the simplest
choice is given by
$$
n_{\mu\nu}=\epsilon_{\mu\nu}
\eqn\twodtwist
$$
The explicit form of the four and higher dimensional twists
can be found in the quoted literature.

We now set aside lattice gauge theory and return to our problem.
The twisted reduced version of (2.3) becomes:
$$
S={1\over 2}\sum_{\mu}Tr(\Gamma_{\mu}\phi\Gamma_{\mu}^{-1}
-\phi)^2+TrV(\phi).
\eqn\rstring
$$
The equivalence of the planar approximation to (2.3) with
\rstring\ follows from earlier work on reduced models
\REF\en{T. Eguchi R. Nakayama, \plb {\bf 122 B} (1983) 59.}
[\tek,\en].  For simplicity we consider the two-dimensional
case.  First, one can always write a hermitean matrix
in the $\Gamma_{\mu}$ basis
$$
A(q)=\Gamma_1^{k_1}\Gamma_2^{k_2}\qquad k_{\mu}=\epsilon_{\mu\nu}
q_{\nu},
$$
with $q_{\nu}$ defined modulo $N$.  We can expand the hermitean
matrix $\phi$ in the $\Gamma$-basis according to
$$
\phi=\sum\phi_q A(q).
$$
Some useful properties of the $A$-matrices are:
$$
A(q)^{\dagger}=A(-q)e^{{2\pi i \over N}\langle k|k\rangle}
\qquad \langle k^i|k^j\rangle =\sum_{\mu<\nu}n_{\mu\nu}
k_{\nu}^ik_{\mu}^j,
$$
and
$$
\Gamma_{\mu}A(q)\Gamma_{\mu}^{\dagger}=e^{2\pi i q_{\mu}/L}A(q) ,
$$
explicitly showing how the $\Gamma$'s implement the lattice
translation group.  Furthermore
$$
TrA(q_1)^{\dagger}A(q_2) = N\delta_L(q_1-q_2),
$$
where the $\delta$-function is defined mod $L$.  Using these
properties, the kinetic term
of our action becomes,
$$
{1\over 2}\sum_{\mu}Tr(\Gamma_{\mu}\phi\Gamma_{\mu}^{-1}
-\phi)^2 +{1\over 2}m^2Tr\phi^2\sim
$$
$$
\sum_q\left[ {1\over 2}m^2 +\sum_{\mu}(1-\cos ({2\pi q_{\mu}
\over L})\right] \phi_q \phi_{-q} e^{-2\pi i \langle k|
k\rangle /N},
\eqn\prop
$$
coinciding with the standard lattice propagator of
a scalar field on a size $L$ $D$-dimensional lattice
up to the last phase factor in \prop.  The role of this
and similar phase factors appearing in the vertices
of the theory is to restore the topological expansion
in large-$N$.  We have now obtained an effective
field theory with scalar variables $\phi_q$.
The phase factors
keep track of the fact that the theory came from
a matrix field theory and that the expansion
in powers of $1/N$ can be represented in terms
of a topological expansion where the leading order
corresponds to spherical topologies, and higher orders
come from surfaces of increasing Euler number.  The phases
cancel completely only for planar graphs,
and for higher topologies the lack of cancellation
leads to suppresions by powers of $1/N^2$. Note
that in \prop\ one shows that the space-time degrees
of freedom of the original theory are coded in the
Fourier transfoms of the ``internal" $SU(N)$ indices.

The construction we just carried out works only
for even dimensions.  For odd-dimensions we can
leave one of the dimensions unreduced, and apply
the reduction prescription to the remaining
even number of dimensions.  This yields an action
\def\onet{{1\over 2}}
$$
S=\int dt \onet Tr\dot\phi(t)^2+\onet\int dt Tr
(\Gamma_{\mu}\phi(t)\Gamma_{\mu}^{-1}
-\phi)^2
$$
$$
+\int dt Tr V(\phi(t)).
$$
We can consider a slightly more general model by
including a general coupling in the hopping term:
$$
-S(\phi)=\sum_{\mu}\onet R^2 Tr\phi\Gamma_{\mu}\phi\Gamma_{\mu}
^{\dagger}-\onet m^2\phi^2-TrV_0(\phi)
$$
$$
V_0(\phi)={g_3\over \sqrt{N}}\phi^3+{g_4\over N}\phi^4+\ldots
\qquad m^2=2DR^2+m_0^2.
\eqn\raction
$$
The hopping term is the first term in $S(\phi)$.  Adding the
effect of world-sheet curvature is easy, we simply
replace the hopping term according to
$$
Tr(\Gamma_{\mu}\phi\Gamma_{\mu}
^{\dagger}-\phi)^2\mapsto
TrA (\Gamma_{\mu}\phi\Gamma_{\mu}^{\dagger}-\phi)
A(\Gamma_{\mu}\phi\Gamma_{\mu}^{\dagger}-\phi)
\eqn\ractioncur
$$
Our proposal to describe strings in dimensions higher than
one is to investigate the properties of \raction\ and
\ractioncur. We have effectively reduced the problem
to a one-matrix model.  The
difficulty lies however in the presence of the twist
matrices.  We begin our analysis in the following section.

\chapter{PROPERTIES OF THE REDUCED STRING ACTION}

Some general features of \raction\ and
\ractioncur\ are easy to
extract.  If the hopping term is
very small, we are in the limit
where the space-time lattice points are
very far apart and the
whole surface is mapped into a single point,
or various points
incoherently.  This is the pure gravity
regime, $\gamma_{\rm st}$
\def\gst{\gamma_{\rm st}}
is expected to be $-1/2$ and the induced metric plays no role.
As the hopping term is increased, it becomes more likely for
the surface to occupy more and more space-time lattice sites,
the induced metric and singular embeddings in principle
begin to play important roles, and we can expect a transition
to a phase different from pure gravity at some critical
value of the hopping coupling $R^2$.  We will show presently
that in the most na\"{\i}ve approximation, we are driven
to a branched polymer phase with $\gst=1/2$,
and at the transition point $\gst=1/3$.  A more careful
analysis will show a much broader spectrum of possibilities.
Note that the role played by $R$ is similar to the compactification
radius in the study of $c=1$ strings propagating in a
circle of radius $R$
\REF\grosskleb{D. Gross and I. Klebanov, \npb {\bf B 344}
(1990) 475.}
[\grosskleb].

To deal with \raction\ we decompose the matrix $\phi$ into
its eigenvalues and angular variables as usual
$$
\phi=U^{-1}\lambda U\qquad \lambda=
{\rm diag}(\lambda_1,\lambda_2,\dots,\lambda_N),
$$
The angular integration produces an effective action
$$
e^{\Gamma_{\rm eff}(\phi)}=\int dU e^{R^2\sum_{\mu}Tr\lambda
\Gamma_{\mu}(U) \lambda \Gamma^{\dagger}(U)}
$$
$$
\Gamma_{\mu}(U)=U\Gamma_{\mu}U^{-1}
$$
$$
Z=\int d\phi e^{\Gamma_{\rm eff}[\phi]-V_0(\phi)}
\eqn\effaction
$$
The simplest representation of $\Gamma_{\rm eff}$ is given
in terms of the leading order expansion in $R^2$:
$$
Z=\int d\phi e^{N\left(R^2{D\over N}(Tr\phi)^2-V_0
(\phi)\right)}.
\eqn\polyaction
$$
The critical properties of \polyaction\ are very similar
to those of the curvature model in [\tata].  What the
first term in $\Gamma_{\rm eff}$ is describing is the
breaking of the surface.  The Feynman graph representation
of a term like $(Tr \phi^n)^2$ is a vertex where two
surfaces osculate at one point, and the coordination
number of the touching
vertices in each of the surfaces is $n$.
If we have instead a term of the form
$(Tr\phi^n)^p$, it represents $p$-surfaces sharing one
point.  When there are no products of traces, the Feynman
graph expansion for the free energy or the connected
correlators involves only the sum over connected orientable
surfaces.  When products of traces are included we
obtain an expansion where we have surfaces of different
topologies touching at points determined by the new
vertices.  The new graphs look like ``daisies".  In
the original curvature model introduced in [\tata]
the initial partition function is
$$
Z=\int d\phi e^{-N\left( \onet \phi A\phi A+g Tr\phi^4
\right)}
\eqn\curmodel
$$
The angle integral is again very difficult to carry out in
this model, and the simplest non-trivial approximation
is to study the action:
$$
S_{\rm eff}=N\left(\onet Tr\phi^2 +gTr\phi^4
+{g'\over N}(Tr\phi^2)^2\right).
$$
For real $A$, the coupling $g'=(TrA^2/N-(TrA)^2/N^2)$
is always positive.  In this case as we will see later
one cannot escape the pure gravity phase.  To leading
order in $1/N$, the graphs have the topology of
spheres touching at one point, where any two spheres
in the graph can share at most one common point, and
no closed loop of spheres can appear.  A generic
graph looks like a tree of spheres.
In the planar approximation it was
shown in [\tata] that we have the following possibilities:

1). $g'>0$, $\gst =-1/2$ and we are
in the pure gravity regime.

2). $g_c<g'<0$.  Again $\gst =-1/2$ and we continue
in the pure gravity case.  The branching is still
unimportant, and the leading contributions to
the free energy come from graphs with few big spheres
and few small branchings.

3).  At $g'=g_c=-9/256$, $\gst=+1/3$ and the number
of branchings become competitive with the number of
smooth spheres.

4).  For $g<g_c$ we obtain $\gst=+1/2$ and the dominating
term is the one with a product of two traces, thus
the notion of a smooth surface disappears and we end
up with a branch polymer phase.

The same analysis can be carried out with our na\"{\i}ve
approximation \polyaction, and we will only note
the differences with respect to the analysis in
[\tata] we have just reviewed.  The basic differences
are:

1).  Our potential is cubic, $V_0=\onet \phi^2+g\phi^3
/\sqrt{N}$, the one-cut solution is asymmetrical, and
the new vertex is just given by $Tr(\phi)^2$.  These
are purely technical differences and nothing in the
analysis changes substantially.

2).  We are always in the $g'<0$ region.  In the
original model in [\tata] we can reach this region
only after analytically continuing the matrix
$A\mapsto iA$ in order to obtain the polymer phase.
However, in this case the interpretation of the matrix
$A$ in terms of local world-sheet curvature loses its
meaning.

Our simplified
model then contains a pure gravity phase, and
intermediate phase with $\gst=1/3$ and a
branched polymer phase with $\gst=1/2$.  We should
like to remark that the exact solution to the
pure curvature model \curmodel\ is known only for
Penner's potential
\REF\makeenko{Yu. Makeenko and L. Chekhov,
{\it The Multicritical Kontsevich-Penner Model}.
NBI-HE-92-03, January 1992.}
[\makeenko], however in this case the full theory
has the symmetry $A\mapsto {\rm const.}A^{-1}$
making the interpretation of
$A$ in terms of world-sheet curvature rather
doubtful.

\chapter{PRELIMINARY ANALYSIS OF THE EFFECTIVE ACTION}

We have learned that the basic problem
is the evaluation of the angular integral
$$
e^{\Gamma_{\rm eff}}=
\int dU e^{R^2\sum_{\mu} Tr \lambda U\Gamma_{\mu}U^{-1}
\lambda U\Gamma_{\mu}^{\dagger}U^{-1}}.
\eqn\basicint
$$
The na\"{\i}ve approximation embodied in \polyaction\
already gave us a polymer phase and a critical
point in-between pure gravity and polymers.  To
understand the behavior of strings in dimensions
higher than one, we need to obtain information
about the general critical properties of the effective
action in \basicint.  Since $R^2$ is a  hopping parameter,
we can write the hopping term as
$
L_{\mu}=\phi \Gamma_{\mu} \phi \Gamma_{\mu}^{\dagger}$
and in analogy with standard lattice analysis, we could
write an approximation to the effective action
in terms of a sum over connected graphs of angular
averages over the hopping terms generating the graph:
$$
\Gamma_{\rm eff} = \sum_{ C,{\rm Connected}}
\langle Tr \prod_{l\in C} L(l)\rangle _U,
$$
where the subscript $U$ indicates that the average is
taken only over the angular variables.  To every connected
graph the average over angles associates a set of product of
traces.  We can interpret these traces as different ways
of fracturing and breaking the loop $C$.  They
describe how the embedded surface is collapsed into
pieces.  If we add a local curvature term (the $A$-matrix),
we will obtain more general operators.  Since we do not have
and explicit way of evaluating \basicint\ and since we
are interested in the critical behavior of \raction\ and
not in the fine details of lattice dynamics, it is worth
exploring the type of universality classes of potentials
with arbitrary numbers of products of traces.  There are
two qualitative classes of potentials, depending on whether
they contain a finite or infinite number of traces.  So far
we have analyzed the case of
finite numbers of traces.  Defining
$$
x_i={1\over N} Tr\phi^{2i},
\eqn\trdef
$$
we take the effective action to be a general function
of an arbitrary, but finite number of $x_i$ variables,
$$
\Gamma_{\rm eff}=V(x_1,x_2,\ldots,x_n)=
V({1\over N} Tr\phi^{2},{1\over N} Tr\phi^{4},\ldots
{1\over N} Tr\phi^{2n}).
\eqn\potdef
$$
Thus we are faced with the analysis of the critical properties
in the planar limit of the action
$$
S(\phi)=N^2 V(x_i)+N TrV_0(\phi)
$$
$$
V_0=\sum_k g_k Tr\phi^{2k}.
\eqn\barbonaction
$$
The world-sheet cosmological constant is related to
the the coupling $g_2=e^{-\mu_B}$.
We have made the simplifying assumption
of restricting our considerations to the case of even
potentials.  This is just a question
of technical simplicity, and
nothing changes if we take general potentials.
the analysis can be carried out for any of the Kazakov
potentials
\REF\kazcrit{V.A. Kazakov, Mod. Phys. Lett. {\bf A4} (1989) 2125.}
[\kazcrit].  The solution of the planar approximation
of \barbonaction\ uses the Hartree-Fock approximation,
whi becomes exact in the planar limit.
We first write down the planar equations for (4.4),
then, we replace the traces \trdef\ by arbitrary
variables $x_i$.  This reduces the problem to the original
pure gravity case studied in [\brezin], and then we fix
the variables $x_i$ self-consistently.  The saddle-point
equations and the Hartree-Fock conditions are given by
$$
\onet \sum_p 2p {\tilde g}_p \lambda^{2k-1}={\rm P.V.}
\int d\mu{\rho(\mu)\over \lambda  - \mu}
$$
$$
{\tilde g}_p= g_p+{\partial V(x)\over \partial x_k}
\eqn\saddle
$$
$$
x_k=\int \rho(\mu) \mu^{2k}.
\eqn\hfcond
$$
$\rho(\mu)$ is the density of eigenvalues.
We will denote by ${\tilde V}$ the potential with the shifted couplings
\saddle.  Following the pure gravity case, we look for
one-cut solutions of \saddle.  This means that the loop
operator
$$
F(p)=\int d\mu {\rho(\mu)\over p-\mu}
$$
is given by
$$
F(p)=\onet {\tilde V}'(p)-M(p)\sqrt{p^2-R},
$$
and the polynomial $M(p)$ is determined by requiring that at
large values of $|p|$, $F(p)\sim 1/p + {\cal O} (1/p^3)$.  Since
$V$ depends only on a finite number of traces, the
potential ${\tilde V}$ is a polynomial, and there are
no ambiguities in the determination of $M$.
Expanding $F(p)$ in powers of $1/p$ we obtain the
expectation values of the traces $x_i$,
$$
F(p)= \sum_{k\ge 0} {x_k\over p^{2k+1}} =
{1\over p}+ {x_1\over p^3} +\ldots \,
\eqn\looop
$$
and the first condition $x_0=1$ is the string equation.  From
the other coefficients we can read off
the Hartree-Fock conditions \hfcond.  The density of eigenvalues
is given by
$$
\rho (\lambda )={1\over 2\pi} M(\lambda)\sqrt{R-\lambda^2}
\qquad \lambda \in [-\sqrt{R},\sqrt{R}]
$$
The dependence of $x_k$ on $R$ is obtained following standard
arguments in the planar limit, and we obtain
$$
x_p(R)=\sum_{k\ge 1}{k^2\over k+p} {2p\choose p}{2k\choose k}
({R\over 4})^{k+p} {\tilde g}_k.
\eqn\xsubp
$$
The string equation is given by \xsubp\ with
$p=0$.  We are interested in the analysis of the
critical points
of (4,4), the computation of string susceptibilities, and
in the computation of correlation functions of
macroscopic loop operators. We present only the results,
the details will appear elshewhere
\REF\abc{L. Alvarez-Gaum\'e, J.L.F. Barb\'on and C. Crnkovic,
in  preparation.}[\abc].  If we denote by $\beta$ the coupling
$g_2$, the string susceptibility is determined from
$$
\chi ={dx_2\over d\beta}={\partial x_2\over \partial \beta}
+{dR\over d\beta} {\partial x_2\over \partial R}.
\eqn\stringsus
$$
This is because the coupling $\beta$ multiplies the operator
$x_2$ in the action. Therefore
$\partial \ln Z/\partial \beta$
is proportional to the expectation
value of $x_2$, and the second
derivative of the logarithm of the partition function is the
string susceptibility.
The dependence of $R$ on $\beta$ is read off from the string
equation $x_0=1$.  In particular,
$$
0={dx_0\over d\beta}={\partial x_0\over \partial \beta}
+ {dR\over d\beta}{\partial x_0\over \partial R}
$$
allows us to read the equation determining the critical
points,
$$
{d\beta \over dR}=-{\partial_R x_0 \over
\partial_{\beta}x_0 },
\eqn\critpoint
$$
Since the denominator in \critpoint\ is proportional to
$R^2$, the critical points are determined by the zeroes
of the numerator.  We can similarly write
$\chi$ according to
$$
\chi =\partial_{\beta} x_2 + \partial_{\beta} x_0
{\partial_R x_2\over \partial_R x_0}
\eqn\newsus
$$
The singularities in the behavior of $\chi$ come
from the ratio of derivatives in \newsus.  For the
standard Kazakov critical points [\kazcrit] \newsus\
does not blow up at the critical point $R_c$, and
$\chi-\chi_c\sim R-R_c$, however at the $m$-th
critical point, $\beta-\beta_c\sim (R-R_c)^m$, and
this yields $\gst =-1/m$.  As we will show presently,
the critical points we describe have $\gst >0$ and $\chi$
will be singular at the critical point.  All
the phase transitions are second order.

In analyzing \saddle, \hfcond\  it is useful to
distinguish between two types of derivatives.  When we
differentiate with respect to $R$, we can consider
first the explicit dependence of the equations without
taking into account the Hartree-Fock condition.  If
we ignore the implicit
dependence on $R$ of $\tilde g_k$
we call this derivative $D_R$, and everything is
as in the pure gravity case. It is not difficult to
show that
\def\ap{{2p\choose p}\left({R\over 4}\right)^p}
\def\ak{{2k\choose k}\left({R\over 4}\right)^k}
\def\al{{2l\choose l}\left({R\over 4}\right)^l}
\def\dr{D_R}
\def\rfour{\left({R\over 4}\right)}
\def\gt{{\tilde g}}
\def\parr{\partial_R}
$$
\dr x_p(R)=\ap \dr x_0(R).
\eqn\derr
$$
Including the $R$ dependence coming from \hfcond\
we have an extra contribution.  The derivatives with
respect to $R$ including the Hartree-Fock induced
dependence on $R$ will be denoted by $\parr$.  Thus,
$$
\parr X_p=\dr x_p +\sum_{k\ge 1} {k^2\over k+p}
{2k\choose k}{2p\choose p}\rfour ^{k+p} \parr \gt
$$
$$
= \dr x_p +\sum_{k\ge 1} {k^2\over k+p}
\ap \ak\sum_q\partial_{pq}V \parr x_q.
\eqn\totalr
$$
Defining the matrix
$$
U''_{pq}=\sum_{k\ge 1} {k^2\over k+p}
\ap \ak \partial_{pq}V,
\eqn\udef
$$
we can compute $\parr x_p$ in terms of
$\dr x_0$:
$$
\parr x_p=[(1-U'')^{-1} A]_p \dr x_0,
\eqn\dofxp
$$
where $A$ is a vector whose $p$-th component
equals $\ap$. When $p=0$, we obtain the criticality
condition implied by (4.10):
$$
\parr x_0=[(1-U'')^{-1} A]_0 \dr x_0=0.
\eqn\critcon
$$
Notice that the criticality condition splits into
two terms. The first one is related to the
function including the couplings with multiple
traces,
$$
[(1-U'')^{-1} A]_0=0,
\eqn\typeone
$$
and the second is equivalent to the criticality
condition for the Kazakov critical points:
$$
\dr x_0=0.
\eqn\typetwo
$$
The term in (4.11) which may lead to singularities
in $\chi$ and to positive $\gst$
is given by
$$
{\parr x_2\over \parr x_0}=
{[(1-U'')^{-1} A]_2\over [(1-U'')^{-1} A]_0}.
\eqn\suscrit
$$
In \critcon\ we can have three possibilities: 1). The
polymer couplings in $U$ become critical but
the gravity part does not.  The zeroes of
$d\beta/dR$ come only from the polymer contribution.
 2). The gravity contribution is the only
one generating zeroes of $d\beta/dR$.  3). Both
terms becomes critical.  In the first case we
have a theory of polymers, and little is remembered
of the coupling to gravity. In the second case
the polymer degrees of freedom are completely
frozen and we reproduce the string susceptibility
and exponents of the Kazakov critical points.  The
third and more interesting case is when both the
``polymer" matter and gravity become critical
simultaneously.  This is the case where we can
find novel behavior.  Notice also that unless we
tune the numerator in \suscrit\ we will generically
obtain $\gst >0$ in cases 1) and 3).  If
\def\rc{R_c}\def\bec{\beta_c}
$\rc,\bec$ are the critical values of
$R$ and $\beta$, and if near the critical point
$$
[(1-U'')^{-1} A]_0\sim (R-\rc)^n\qquad
\dr x_0\sim (R-\rc)^m,
\eqn\nmcrit
$$
then
$$
\beta-\bec\sim (R-\rc)^{n+m+1}.
\eqn\bofrcrit
$$
If we do not tune the numerator of \suscrit\ to
partially cancel the zeroes in the denominator, we obtain
$$
\chi\sim{1\over (R-\rc)^n}.
$$
The previous two equation give a string susceptibility
for the $(n,m)$-critical point:
$$
\gst={n\over n+m+1}.
\eqn\nmstring
$$
By tuning the numerator in \suscrit\ we can change
the numerator in \nmstring\ to any postive integer
smaller than $n$
$$
\gst={p\over n+m+1}.
$$
We will show below that only in the case $n=1$ the model
maintain some resemblance with the properties one
would expect of a non-critical string.  We can
define operators creating macroscopic loop only
in that case.  For $n>1$ the polymerization of the
surface is so strong that there is no room left
to open macroscopic loops.  The simplest case to
study is the one where the function $V$ depends
on a single trace $V=V(x_l)$.  In this case the
matrix $U''$ becomes:
$$
U''_{pl}={l^2\over l+p}\ap\al V''\qquad
p=1,2,\ldots.
$$
The matrix $U''$ contains a single non-vanishing
column in position $l$.  The inverse of $1-U''$ is:
$$
(1-U'')^{-1}= 1+{U''\over 1-{l\over 2}(\al)^2 V''},
$$
therefore,
$$
[(1-U'')^{-1}A]_p=
$$
$$
\ap +
{1\over 1-{l\over 2}(\al)^2 V''}{l^2\over p+l}
(\al)^2\ap V''
$$
$$
[(1-U'')^{-1}A]_0=
{1+{l\over 2}(\al)^2V''\over
1-{l\over 2}(\al)^2V''}.
\eqn\xpxo
$$
{}From these formulae we can immediately derive
the string susceptibility $\chi$.  Without loss
of generality we can restrict our considerations
to the case when $l=2$.  In this case many of the
equations simplify and all the main results remain
the same.  The one-point functions $x_0,x_2$ are
given by
$$
1=x_0(R)= \sum_{k\ge 1} k \ak g_k+
12 V'(x_2)\rfour^2
$$
$$
x_2= \sum_{k\ge 1}{6 k^2\over k+2} {2k\choose k}
\rfour^{k+2} g_k +36 V'(x_2)\rfour^4.
\eqn\xtxo
$$
Restricting the general formulae derived before to
this case we obtain:
$$
x_2=3\rfour^2+3\sum_{k\ge 1}{k(k-2)\over k+2}
{2k\choose k}\rfour^{k+2} g_k
$$
$$
\parr x_2={6\rfour^2\over 1-36\rfour^4 V''}
\dr x_0
$$
$$
R\dr x_0 =2+\sum_{k\ge 1}k(k-2)
{2k\choose k}\rfour^{k} g_k.
\eqn\simplemod
$$
The critical points come from the vanishing of the
last equation and from
$$
1+36\rfour^4 V''=0.
$$
An argument similar to the one presented at the end of
[\tata] shows that the critical point for the transition
between the polymer and the pure gravity phases
is generically second order, although by tuning
of the parameters it could be made of higher order.  Hence
we can define a continuum limit for the critical points
\nmstring.

To explore the new critical points in some detail, we
compute the correlation functions of macroscopic
loop operators.  The simplest loops to compute are
the ones analogous to those appearing in pure gravity
(see for instance
\REF\mss{G. Moore, N. Seiberg and M. Staudacher, \npb
{\bf B 362} (1991) 665.}[\mss]).  They are obtained by taking
the limit $k\rightarrow\infty$ of $Tr\phi^{2k}$.  In
terms of the original reduced model these are loop on
the surface whose position is averaged over the target
space.  One could also define loops with definite
positions in the target space.  The expectation values
of the latter would certainly give crucial information
on the reduced formulation of string theory, but we
have not yet calculated them.  For the simpler, pure
gravity loops of length $l$ we take the definition
$$
w(l)=\lim_{k\rightarrow\infty} \sqrt{{\pi\over l}}
\langle Tr \phi^{2k}\rangle =
N\lim_{k\rightarrow\infty}\sqrt{{\pi\over l}} x_k.
\eqn\lloop
$$
As a matter of convenience we choose the critical values
for $\beta$ and $R$ as $\bec=-1, R_c=1$.  In the
Kazakov critical points, $\gst=-1/m$ and the scaling
limit is taken to be $Na^{2-\gst}=\kappa^{-1}$.  It is
a simple but non-trivial fact that the correct way to
define a loop of length $l$ in this case is to scale
$k$ according to $ka^{-2\gst}=l$ as $k\rightarrow\infty$
(see for example [\mss] and references therein).  In our
case we will restrict for simplicity
our computations to the case of
potentials $V=V(x_2)$.  The relevant formulae to be
used are (4.23-25).  We have in general that
\def\parb{\partial_{\beta}}
$$
{dx_k\over d\beta}=\parb x_k-\parb x_0{\parr x_k
\over \parr x_0},
\eqn\dxdb
$$
and from
$$
\parb x_k={24\over k+2}{2k\choose k} \rfour^{k+2}
\left( 1+V''\parb x_2\right)
$$
we obtain
$$
\parb x_k={2\over k+2}\ak \parb x_0.
\eqn\relone
$$
Simple manipulations yield:
$$
{\parr x_k\over \parr x_0} =\ak
{1+36{2-k\over 2+k}\rfour^4 V''\over
 1+36\rfour^4 V''}.
\eqn\reltwo
$$
and after some algebra we obtain:
$$
{dx_k\over d\beta}={12k\over k+2}{2k \choose k}
\rfour^{k+2}{1\over 1+36\rfour^4 V''}
\eqn\basicl
$$
Up to irrelevant numerical factors this answer
coincides with the one-loop expectation value
in the Kazakov critical points
when $V''=0$.  To go to the continuum limit we have
to choose the scaling variables.  Recall that in the
double scaling limit [\doubles] the free energy
takes the form
$$
F=\sum_{g\ge 0}N^{2-2g} (\mu-\mu_c)^{(2-\gst)(1-g)} F_g
$$
where $\mu=-\log\beta$, and $g$ is the genus of the
triangulations summed over.  The renormalized cosmological
constant is introduced according to
$\mu-\mu_c=a^2 t$ where $a$ is the microscopic lattice
spacing.  Then the effective string coupling constant
is defined by $N a^{2-\gst}=\kappa^{-1}$.  Hence the
scaling variable associated with the string susceptibility
is $\chi = a^{-2\gst} u$.  At the $(n,m)$-critical point
$\chi$ behaves as $\chi\sim(1-R)^{-n}$.  Thus the
scaling variables are
$$
\chi\sim{1\over (1-R)^n}=a^{-2\gst}u\qquad
1-R=a^{2\gst/n} u^{-1/n}
$$
$$
\beta-\beta_c=\beta+1=a^2 t.
\eqn\scalevar
$$
For an $(n,m)$-critical point we have
$$
1-R=a^{{2\over n+m+1}}u^{{-1\over n}}\qquad
\chi = u a^{{2n\over n+m+1}}\qquad
N a^{2-{n\over n+m+1}}=\kappa^{-1}.
\eqn\critscaling
$$
Let us start with the critical points $(1,m)$.  From
\basicl, and \critscaling\ we easily obtain as
$k\rightarrow\infty$
$$
{d\over dt}\langle Tr\phi^{2k}\rangle=
-{3\over 16}{N a^{2-{2\over m+2}}\over \sqrt{\pi k}}
u e^{-ka^{{2\over m+2}}u^{-1}},
\eqn\prevloop
$$
thus, choosing $ka^{2/m+2}=l$ fixed as $k\rightarrow\infty$
and $a\rightarrow 0$ together with \critscaling\ we obtain
a finite result
$$
{d\over dt}w(l)=-{3u\over 16\kappa l} e^{-l/u}.
\eqn\oneloop
$$
If we take instead $n>1$, it is easy to show that
there is no possible definition of the macroscopic
length $l$ giving a finite result for \prevloop.  When
$n>1$ we have
$$
 1+36\rfour^4 V''\sim (1-R)^n.
$$
Repeating previous arguments we arrive at
$$
{d\over dt}\langle Tr\phi^{2k}\rangle=
-{3\over 4}Na^{2-{2n\over m+n+1}}{1\over \sqrt{\pi k}}
e^{-ka^{{2\over m+n+1}}u^{-1/n}},
$$
and there is no way of choosing the scaling behavior of
$k$ giving a finite value without extraneous renormalization
of the loop operator.  Hence in the phases with $n>1$ the
polymer couplings dominate completely the critical behavior
in spite of the fact that gravity becomes also critical
in such a way that there is not enough room to open
macroscopic loops.  In the case of $n=1$ we can also
calculate two- and multiloop correlation functions.  The
result obtained for the two-loop correlators is rather
suggestive,
$$
w(l_1,l_2)={e^{-(l_1+l_2)/u}\over l_1+l_2}+
{u\over l_1l_2} e^{-(l_1+l_2)/u}
\eqn\twoloop
$$
and similar results for multiloop correlators.  The lesson
we can draw from \twoloop\ is that in the simple case of
$n=1$ the effect of the polymer couplings is to contribute
an extra state (the last term in \twoloop\ ) which resembles
very much the contribution one would expect of a tachyon.
This term represents the breaking of the cylinder
interpolating between the two loop of lengths $l_1,l_2$
into two disks osculating at one point.  We can represent
the last term in \twoloop\ as
$$
\langle w(l_1)P\rangle {1\over \langle PP\rangle}
\langle P w(l_1)\rangle
$$
where $P$ is the punture operator.  This pattern
repeats in the higher loop planar correlators, and it is
reminiscent of the residue of the tachyon pole in the
Belavin-Knizhnik theorem
\REF\bk{A.A. Belavin and V. Knizhnik, \plb {\bf 168 B}(1986)
201; J.B. Bost and T. Jolicoeur, \plb {\bf 174 B}(1986) 273;
R. Catenacci, M. Cornalba, M. Martellini and C. Reina,\plb
{\bf 172 B}(1986) 328.}
[\bk].  This interpretation, though tempting has to be
taken with several grains of salt because in the types
of effective actions we are studying, it is hard
to see any dependence on dimensionality.  Nervertheless
we find this result as
encouraging evidence that our approximation captures
some of the expected properties of non-critical strings
for $D>1$.

The form of the two-loop operator \twoloop\ does not
depend on the form of the potential $V(x_1,\ldots x_n)$
as long as we consider $(1,m)$-critical points.  Hence
only the quadratic part of $V$ determines the critical
properties of macroscopic loops.  Whether these conclusions
will still hold when we consider potentials $V$ depending
on an infinite number of traces is currently under
investigation [\abc].

To conclude this long section we mention that we can
study at least in part the spectrum of scaling operators
by perturbing the criticality conditions.  The string
equation is then modified according to
$$
1+\beta=(1-R)^{n+m+1}+\sum t_k^B (1-R)^{k+1}
$$
The subleading contributions depend on both the polymer
and Kazakov perturbations of the critical conditions
\critcon. The bare couplings of the scaling operators
are the $t_k^B$ parameters.  Using \critscaling\ the string
equation becomes
$$
t=u^{-{n+m+1\over n}}+\sum_k t_k^R u^{-{k+1\over n}},
\eqn\stringeq
$$
and the $t_k$ are the renormalized couplings.  In the
original model these couplings would hardly exhaust the
scaling operators, although they may represent a
significant subset associated to the
simplest loop operators.  If we call $\sigma_k$ the
scaling operator associated to the coupling $t_k$, its planar
correlators at the $(n,m)$-critical point follow from
\stringeq\
$$
{du\over dt_k}=-u^{-{k+1\over n}}{du\over dt}
$$
and
$$
\langle \sigma_k PP\rangle={n\over n+m+1}
t^{{k-m-2n\over n+m+1}}.
$$
This concludes our brief study of the effective actions
depending on an arbitrary, but finite number of couplings.

\chapter{CONCLUSIONS AND OUTLOOK}

We have presented a very preliminary analysis of a
new way of formulating strings theories in dimensions
higher than one.  Our approach is a direct application
of the reduced formulation of large-$N$ field theories
[\ek,\tek].  We believe that it is worth while exploring
further the properties of these reduced models to gain
insight into the behavior of strings beyond $D=1$.

We have presented the general analysis of effective actions
in the planar limit containing arbitrary, but finite
numbers of traces \barbonaction.  We were able to show
that beyond some critical coupling, $\gst$ becomes positive
and takes values of the form $\gst=n/n+m+1$ for any positive
integers $n,m$.  We should continue exploring the
properties of these critical points and their scaling
operators.  Obviously the more challenging case of studying
arbitrary functions $V$ depending on an infinite number
of traces is crucial before we can draw any conclusions
concerning the properties of our model.  This together
with the study of general properties of \basicint\
should shed further light into the properties of non-critical
strings in interesting dimensions $D=2,3,4$.  It is
very important to obtain in what way the effective actions
studied in the previous sections depend on dimensionality.

To understand better the transition between $\gst <0$ and
$\gst >0$ in the models analyzed, we calculated
the correlation functions of
several macroscopic loops of
lengths $l_1,l_2,\ldots$.  We found that only for
critical points of the form $(1,m)$ is it possible
to open macrocopic loop on the surface.  For $n>1$
the polymer couplings dominate the continuum limit
and there is no room to open such loops.  Furthermore
in the case of $n=1$ we encountered a new state
which represents the breaking
of the surface into two pieces
touching only at one point, qualitatively in agreement with
what one would expect from the presence of a tachyon
beyond $D=1$.  Whether this conclusion will remain
after one includes more realistic effective actions
remains to be seen.  However this ``stringy" aspect
of our effective action seems encouraging.
We can also compute in the reduced models the explicit
form of ordinary vertex operators which describe embedded
surfaces passing through specified
target space points.  Correlation
functions of vertex operators should
allow us to better assess
the quantum geometry of the
phases for $\gst >0$.  Furthermore,
the study of vertex operators should also help
in the interpertation of the operators
$\prod_nTr\phi^n$ in terms of processes tearing up the
embedded surface and the effect of tachyons.  Finally
it may also be useful to analyze strings
in $D>1$ numerically using
the reduced actions.

Mnay of these issues are currently
under investigation.
Details concerning the topics covered in this lecture
together with further results will appear elsewhere
[\abc].

{\bf ACKNOWLEDGEMENTS}.  One of us (L. A.-G.) would like to
thank the organizers of the Trieste Spring Workshop 1992 for
the opportunity to present these results in such a stimulating
environment.

\endpage
\refout
\end